# From Perturbation Theory to Confinement: How the String Tension is built up [*] [†]

M. García Pérez [a], A. González-Arroyo [b] and P. Martínez [b]

[a] Instituut-Lorentz for Theoretical Physics,
University of Leiden, PO Box 9506,
NL-2300 RA Leiden, The Netherlands.

[b] Departamento de Física Teórica C-XI,
Universidad Autónoma de Madrid,
28049 Madrid, Spain.

We study the spatial volume dependence of electric flux energies for SU(2) Yang-Mills fields on the torus with twisted boundary conditions. The results approach smoothly the rotational invariant Confinement regime. The would-be string tension is very close to the infinite volume result already for volumes of $(1.2 \text{ fm.})^3$. We speculate on the consequences of our result for the Confinement mechanism.

## 1. Introduction

An interesting approach towards the investigation and understanding of Yang-Mills field theory is the study of the dependence of the system on the size of the spatial volume where the theory is defined. For small volumes the behaviour of Yang-Mills fields is well described by Perturbation Theory. For large volumes the Physics becomes volume independent and we obtain the usual Confinement phase. By studying the transition between these two limits one hopes to be able to disentangle part of the structure of the QCD vacuum and/or develop semi-quantitative analytical calculational techniques. A priori, many possibilities are conceivable concerning the big to small volume transition: abrupt or smooth behaviour, one or several scales. The results that we are reporting here are intended to provide answers to these questions.

The previous program was initiated by M. Lüscher [1], who employed a Hamiltonian formulation to study the system on a small box and with purely periodic boundary conditions. Much work was done later in this direction both analytically and numerically ( for a review see Ref. [2]). Our line of attack differs from that of Lüscher by our choice of twisted boundary conditions [3] in space. These b.c. respect periodicity for gauge invariant quantities and lead to a very different dynamics at small volumes. Our choice of the twist vector for SU(2) $\mathbf{m} = (1,1,1)$, is together with the purely periodic case the only one preserving the cubic symmetry of the box. Although for large volumes the dependence on boundary conditions drops out, some boundary conditions might seed the way to the large volume vacuum in a more efficient way than others ( think about twist in the 1-dimensional Ising model as the simplest example). In any case, our study will serve to complement the picture obtained with purely periodic boundary conditions.

The results that we are presenting here are based on Monte Carlo simulations performed on lattices $N_s^3 \times N_t$ with the afore-mentioned twisted boundary conditions. Previous results with these b.c.'s were presented in Ref. [4,5]. In this work we have concentrated in measuring the ground state energies of the different electric-flux sectors.

On the torus there is a $Z(2)^3$ symmetry whose representations can be labelled by a 3-vector of integers (modulo 2) **e**. This quantity can be interpreted as a Z(2) conserved electric-flux vector [3]. The corresponding ground state energy in each sector will be noted $E(\mathbf{e})$. In the Confinement phase these energies grow with respect to the vacuum energy linearly with the size of the box ($l_s = N_s.a$, $a$ the lattice spacing) with a coefficient proportional to the string tension. If we introduce the following quantities

$$\Sigma_n \equiv \frac{(E(\mathbf{e}) - E(\mathbf{0}))}{l_s \times \sqrt{n}} \qquad (1)$$

where $n = e_1 + e_2 + e_3$ takes the values 1,2 and 3, the prediction is that for large sizes $l_s$ the three $\Sigma$ go to a constant equal to the string tension. For small volumes, however, these quantities behave very differently. $\Sigma_1$ and $\Sigma_2$ receive contributions from Perturbation Theory. To leading order they should behave as $\sqrt{2}\pi/(l_s^2 \sqrt{n})$ for n equal 1 and 2 respectively. $\Sigma_3$ is zero to all orders in P.T. and its leading contribution for small sizes $l_s$ comes through tunneling via a $Q = \frac{1}{2}$ instanton [7] contribution. This is clearly seen in Ref. [5] were Monte Carlo simulations for this quantity are presented and compared with the prediction of the dilute gas approximation. The present work can be considered as an extension of that paper to larger torus sizes and inclusion of the other two $\Sigma$ values.

---

[*] Based on the talk given by A. González-Arroyo at Lattice'93 (Dallas, 12-16 Oct, 1993)
[†] Partially financed by CICYT; M.G.P. supported by HC & M EC fellowship

## 2. Results

The following results have been obtained by simulations performed on our transputer-based machine RTN. The details of the simulations are the same as those of Ref. [5]. We have explored a region of $\beta$ values ranging from $\beta = 2.3$ up to $\beta = 2.6$ and $N_s = 4, 6$ and $8$ with $N_t = 128, 128$ and 64 respectively. In order to measure the electric-flux energies we have studied the time dependence of correlators between operators with the appropriate quantum numbers. Essentially, linear combinations of Polyakov loops with spatial winding numbers (modulo 2) equal to the corresponding electric flux value. The choice of the appropriate linear combinations is crucial to enhance the signal over the noise. We have employed variants of the by now standard fuzzying and smearing algorithms. Details of the operators and comparison of different possibilities will be given elsewhere [8].

In order to present our results we will give the values of the $\Sigma_n$ in physical units as a function of the length of the torus in fermis. To convert lattice to physical units we use the phenomenological formula

$$a(\beta) = 400 \, \exp(-\frac{\log 2}{0.205}\beta) \text{ fm}. \qquad (2)$$

which accounts for scaling in the string tension and the deconfinement temperature in the corresponding region of $\beta$ values. The absolute scale is fixed by the ad-hoc requirement that the infinite volume string tension $\sigma$ is equal to 5 fm$^{-2}$. Given the previous formula our choice of $\beta$ values has been guided by the will to obtain the same values of $l_s$ for different values of $N_s$.

In Fig. 1 we present our results. The three values of $\Sigma$ are plotted and seen to approach each other for large $l_s$. At small sizes the quantities in question follow the prediction of the weak coupling approximation given by the solid curves. Each curve represents the sum of a perturbative part computed in Ref. [6] plus a semiclassical dilute gas of $Q = \frac{1}{2}$ instantons contribution. The absolute magnitude of the latter piece is the only unknown quantity which has been fitted to the data. In all cases the coefficients are allowed to depend on $N_s$ and hence there are 2 curves ($N_s = 4$ and 6) for $\Sigma_{1,2}$ and three curves ($N_s = 4, 6, 8$) for $\Sigma_3$. Scaling is nevertheless very approximately true both for the data points and the curves.

Beyond $l_s = 0.7$ fm the dilute gas approximation breaks down and we did not extrapolate our curves much beyond that region. The data, however, behave smoothly beyond this point and at $l_s = 1.2$ fm the three $\Sigma$ quantities have overlapping error bars, yielding a common estimate consistent with the value of 5 fm$^{-2}$ of the infinite volume point. Hence, within errors at this size rotational invariance has been recovered and we are quantitatively (within a 10 % accuracy) close to the infinite volume limit. Our results are consistent with the ones obtained in Ref. [4] which are also plotted.

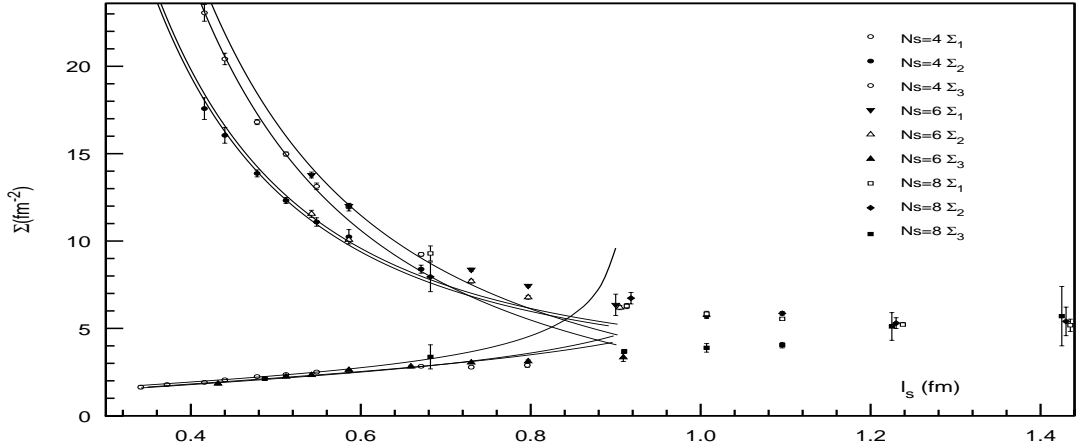

Figure 1. We plot $\Sigma_1$, $\Sigma_2$ and $\Sigma_3$ in fm$^{-2}$ as a function of $l_s$ in fm.

## 3. Conclusions

The whole picture that emerges from our results is as follows. Up to $l_s \approx 0.7$ fm the would-be string tension $\Sigma_n$ is well described by the weak coupling expansion. At the border of this region the values of $\Sigma_n$ are 3., 7.5 and 8.2 fm$^{-2}$ for decreasing $n$. Beyond that point we do not have analytical control, but the data show that these quantities approach smoothly the infinite volume value. What can we learn from these facts? It is tempting to speculate that the $Q = \frac{1}{2}$ instanton is related to the structure of the large volume limit responsible for Confinement. To clarify what we have in mind you must realize that the torus-instanton can be looked at as a periodic solution in infinite space. Even on a torus there is a whole family of related classical configurations with periodicity a fraction of the torus size. They are a lattice of closely packed $Q = \frac{1}{2}$ lumps. Volume independent behaviour can be obtained if the lump-lattice spacing stays constant in physical units. Lattice estimates of the SU(2) topological susceptibility tell us that this spacing has to be $l_s \approx 0.7$ fm !!. The presence of a lattice of lumps breaks rotational invariance but it is conceivable that it costs very little free energy to disorder the system. Indeed if on a torus you have k $Q = \frac{1}{2}$ lumps arranged as a lattice, the index theorem tells us that there is a $4k$ dimensional family of self-dual configurations with the same action ($4k$ are the degrees of freedom of a gas of k particles in 4-space). Furthermore, if we make an estimate of the string tension for such a gas we get roughly the right value. We hope to have convinced you that a picture of the QCD vacuum as a gas of lumps of topological charge $\frac{1}{2}$ might not be that crazy (See also Zhitnitskys talk in these Proceedings). We will postpone to Ref. [8] a more detailed explanation of these ideas.